\documentclass[12pt]{article}
\newcommand{\1}{\begin{equation}}
\newcommand{\2}{\end{equation}}
\newcommand{\ea}{\begin{eqnarray}}
\newcommand{\ee}{\end{eqnarray}}
\newcommand{\bee}{\begin{eqnarray*}}
\newcommand{\eee}{\end{eqnarray*}}

\newcommand{\dnd}[2]{{\frac{{\rm d}#2}{{\rm d}#1}}}

\newcommand{\pd}[2]{{\frac{{\partial}#2}{{\partial}#1}}}

\newcommand{\pdnd}[2]{{\frac{\partial #2}{\partial #1}}}

\newcommand{\op}[1]{\hat{#1}}
\newcommand{\erw}[1]{\left\langle\, #1\,\right\rangle}

\newcommand{\ke}[1]{\,|\,#1\,\rangle}

\newcommand{\pp}{-\!\!\!\!\!\!}

\newcommand{\de}{{\!\rm d}}
\newcommand{\e}{{\rm e}}

\newcommand{\g}{{\!\,=\,\!}}

\newcommand{\ii}{{\rm i}}

\newcommand{\sa}{\left[ \begin{array} {c} }
\newcommand{\se}{\end{array}\right]}

\usepackage{texdraw}
\usepackage{graphicx}
\begin{document}

\begin{center}
{\Large\bf Dynamics of Uniform Quantum Gases, I: Density and Current Correlations}\\*[8mm] {\large J. Bosse}\\*[1.5mm]
{\small Institute of Theoretical Physics, Freie Universit{\"a}t,
Berlin 14195, Germany}\\*[3mm] {\large K. N. Pathak}\\*[1.5mm]
{\small Department of Physics, Panjab University, Chandigarh 160
014, India}\\*[3mm] {\large G. S. Singh}\\*[1.5mm] {\small
Department of Physics, Indian Institute of Technology, Roorkee 247
667, India\\*[3mm]
(15 December 2009 )}\\
*[1cm]

\end{center}

\begin{abstract}

A unified approach valid for any wavenumber $q$, frequency $\omega$, and temperature $T$ is presented for uniform ideal quantum gases allowing for  a comprehensive study of number density and particle--current density response functions. {\em Exact analytical} expressions are obtained for spectral functions  in terms of polylogarithms.  Also, particle--number and particle--current static susceptibilities are  presented which, for fugacity less than unity, additionally involve Kummer functions.  The  $q$-- and $T$-- dependent  transverse--current static susceptibility is used to show explicitly that current correlations are of a {\em long} range in a  Bose--condensed uniform ideal gas but for bosons at $T> T_c$ and for Fermi and Boltzmann gases at all temperatures these correlations are of {\em short} range. Contact repulsive interactions for systems of {\em neutral} quantum particles are considered within the random phase approximation. The expressions for particle--number and transverse--current susceptibilities are utilized to discuss the existence or nonexistence of superfluidity in the systems under consideration.
\\*[5mm]
\\*[5mm]

\end{abstract}

\section{Introduction}
\label{Introduction}

Momentum correlations of infinite range are considered to be a fundamental property of a superfluid \cite{hom:65,for:75}. The complexity in evaluating the relevant finite-temperature expressions to investigate this property in any physical system has probably kept the suggestion almost dormant. For a one-component fluid, the knowledge of longitudinal and transverse particle-current static susceptibilities, $\tilde{\chi}_{\parallel}(q)$ and $\tilde{\chi}_{\perp}(q)$, in the long-wavelength ($q\to0$) limit is required to decide about the existence of long--range correlations.
Since $\tilde{\chi}_{\parallel}(q)$ is exactly known  for any wavenumber $q$ from the famous $f$-sum rule \cite {pin:99},  the study of the property here boils down to the evaluation of $\tilde{\chi}_{\perp}(q)$ for $q\to0$. We would like to investigate nontrivial systems for which the relevant susceptibilities can be determined in exact analytical forms.

If one considers weakly interacting quantum systems and takes into account the interactions within the random phase approximation (RPA), the susceptibilities can be determined from knowledge of the corresponding expressions for a noninteracting system.
Of course, for an ideal Fermi gas at $T\g0$, the longitudinal response functions are well known, see e.g. \cite[Chap. 12]{few:71}.
For a Bose gas, Pines and Nozi{\`e}res \cite[Chap. 4.2, Vol.II]{pin:99} have discussed that $\tilde{\chi}_{\perp}(q)\g0$ at $T\g0$, but it has been emphasized  by Pitaevskii and Stringari \cite[p.97]{pis:03} in the context of an ideal Bose gas  that  evaluation of ($T$--dependent) $\tilde{\chi}_{\perp}(q)$  cannot be carried out so easily as that of $\tilde{\chi}_{\parallel}(q)$. Density correlations at finite $T$  in ideal Fermi and Bose gases have been discussed in Ref. \cite{bae:71}. It is worth mentioning, however, that although transverse--current correlations at finite $T$ are available in semi--analytical and/or numerical forms for  Fermi gases \cite{khg:76, ktp:02} but their neat analytical forms are still lacking.

The main purpose of this work is to present in a comprehensive manner a unified study of response functions of longitudinal and transverse particle--current density (as well as of number density) for all values of $q$, $\omega$ and $T$  for gases obeying Bose--Einstein (BE), Fermi--Dirac (FD), and Maxwell--Boltzmann (MB) statistics. {\em Exact analytical} forms are  obtained for spectral functions as well as  static susceptibilities. These dynamical correlation functions are shown to be expressible in terms of polylogarithms; earlier investigations \cite[and references therein]{lee:95} in terms of polylogarithms have been mainly for thermodynamic quantities. The expressions obtained by us would be useful in  theoretical investigations of many physical properties  enunciated in \cite{ktp:02} and are utilized in the companion article \cite{bps.b:09}, hereafter referred to as Paper II, to study the magnetic susceptibility in quantum gases of {\em charged} particles.
Moreover, there have been interesting new developments in recent years \cite{pis:03,dgp:99, leg:01, pes:08, bdz:08} in the study of low--density atomic quantum gases wherein the particles have short--range interactions and our results might turn out to be useful in
situations where density and/or current correlations become accessible in  experiments on such systems.

The second purpose of this work is to utilize a $q$-- and $T$--dependent expression for $\tilde{\chi}_{\perp}(q)$ to establish the relation  $\lim_{q\to0}\tilde{\chi}_{\perp}(q)\le \lim_{q\to0} \tilde{\chi}_{\parallel}(q)$ from which the existence or nonexistence of long--range correlations is deduced. It is found that while FD and MB gases possess correlations of finite range, and thereby constitute normal fluids, at all temperatures, the BE gas is a normal fluid for $T>T_c$ only. For $T<T_c$, however, the BE gas has current correlations of infinite range even in the noninteracting case implying that the existence of long-range momentum correlations is only a necessary condition for a system to be superfluid.

The outline of the paper is as follows.
In Sec. \ref{basics}, we introduce the definitions of various physical quantities required for our studies and briefly describe the procedure for their evaluation.
In Sec. \ref{current-spectra}, exact analytical expressions for spectral functions are obtained and the results for the current response spectra are presented graphically. The normalized current relaxation spectra are also plotted in this section.
The particle--number and particle--current static susceptibilities are presented in Sec. \ref{Static Susceptibilities}, both numerically and analytically. These results are then applied in Sec. \ref{Condensation and Superfluidity} to examine the issue of Bose condensation and superfluidity in noninteracting as well as interacting systems wherein the influence of interactions is considered within RPA. We summarize our findings in Sec. \ref{Conclusion} and some necessary details of our calculations are provided in the Appendix.

\section{Basic Considerations}
\label{basics}

If the interactions between particles are negligible, the
calculation of the number--density response function
\1
\label{def-density-response} \chi(q,t)=\frac{1}{V\hbar}\erw{\left[\op{N}_{\bf q}(t),\op{N}_{\bf
q}^\dagger(0)\right]}
\2
with the number--density operator
$\op{N}_{\bf q}\g\sum_{{\bf k}\sigma} a_{{\bf
k},\,\sigma}^\dagger a_{{\bf k}+{\bf q},\,\sigma}$ as well as of
the particle--current response tensor
\1
\label{def-current-density} \chi_{\alpha\alpha'}({\bf
q},t)=\frac{1}{V\hbar}\erw{[\op{J}^\alpha_{\bf q}(t),\op
{J}^{\alpha'}_{\bf q}(0)^\dagger]}
\2
with the particle--current density
operator $\op{J}^\alpha_{\bf q}=\sum_{{\bf
k}\sigma}\hbar/m\,\left(k_\alpha+q_\alpha/2\right)\, a_{{\bf
k},\,\sigma}^\dagger a_{{\bf k}+{\bf q},\,\sigma}$
($\alpha,\alpha' = x,y,z$) for a system of ${\cal N}$ identical
particles contained in a box of volume $V$ will simplify
considerably. Here $\op{N}_{\bf q}(t)$ and ${\op{\bf J}}_{\bf q}(t)$
are the Heisenberg operators corresponding to $\op{N}_{\bf
q}\g\int\de^3r~\e^{-\ii{\bf q}\cdot{\bf r}}\op{N}({\bf r})$ and
$\op{\bf J}_{\bf q}\g\int\de^3r~\e^{-\ii{\bf q}\cdot{\bf r}}\op{\bf
J}({\bf r})$, respectively, $ a_{{\bf k}\sigma}^\dagger
( a_{{\bf k}\sigma})$ are the Fock-space operators which will
create (annihilate) a particle in the state $\ke{\bf{k}\sigma}$, and
the angular bracket $\erw{\dots}$ denotes thermal average over
the grand canonical ensemble, implying $\erw{\op{N}_{{\bf q}\g0}}\g
{\cal N}$.

The many--particle Fock space averages in
Eqs.(\ref{def-density-response}) and (\ref{def-current-density})
will reduce to averages in the single--particle Hilbert space for a
system of noninteracting particles. In particular, for a uniform ideal quantum gas with
Hamiltonian $\op H\g \sum_{{\bf k}\sigma}\varepsilon_{\bf
k}\, a_{{\bf k}\sigma}^\dagger  a_{{\bf k}\sigma}$ and
$\varepsilon_{\bf k}=\hbar^2 k^2/(2m)$, it is straightforward to show
that the above response functions of number density and particle--current densities
will reduce to
\ea
\label{FNIP-chi}
\chi(q, t)&=&\frac{2n}{\ii\hbar}\sum_{\bf
k}C_{\bf k}\,\sin\left[\frac{\Delta_{\bf k}({\bf
q})}{\hbar}\,t\right],
\ee
which depends on  $q\g|{\bf q}|$ only, and \ea \label{FNIP-chi-alpha-beta}
\chi_{\alpha\alpha'}({\bf q},t)&=& \frac{2n}{\ii\hbar}\sum_{\bf
k}C_{\bf k}\, v_{\bf k}^\alpha({\bf q})v_{\bf k}^{\alpha'}({\bf
q})\, \sin\left[\frac{\Delta_{\bf k}({\bf q})}{\hbar}\,t\right]\;.
\ee
Here $n\g{\cal N}/V$ is the overall number density, $C_{\bf
k}\g\cal {N}_{\bf k}/ \cal N$ denotes the thermal-average fraction
of particles having momentum $\hbar{\bf k}$, $v_{\bf k}^\alpha({\bf
q}) = (\hbar /m)\left(k_\alpha + q_{\alpha} /2\right)$ and
$\Delta_{\bf k}({\bf q}) = \varepsilon_{{\bf k}+{\bf
q}}-\varepsilon_{\bf k}$. Also, ${\cal N}_{\bf
k}=\sum_\sigma\erw{ a^\dagger_{{\bf k}\sigma} a_{{\bf
k}\sigma}}$ and $\sum_{\bf k}{\cal N}_{\bf k}={\cal N}$.

For the {\em uniform} (homogeneous and isotropic) systems studied
here,
\1
\label{chi-LT-tensor}
\chi_{\alpha\alpha'}({\bf
q},t)=\chi_\parallel(q,t)\frac{q_\alpha q_{\alpha'}}{q^2}
+\chi_\perp(q,t)\left(\delta_{\alpha\alpha'}-\frac{q_\alpha
q_{\alpha'}}{q^2}\right)
\2
will have only two independent components, namely the longitudinal and the transverse. These components are given from
Eq.(\ref{FNIP-chi-alpha-beta}) in conjunction with
Eq.(\ref{chi-LT-tensor}) as
\ea
\label{chi-L-of-qt}
\chi_\parallel(q,t)&=&\sum_{\alpha,\alpha'}\frac{q_\alpha
q_{\alpha'}}{q^2}\chi_{\alpha\alpha'}({\bf
q},t)~=~-\frac{1}{q^2}\,\pdnd{t^2}{^2}\chi(q,t),
\ee
where the relation to $\chi(q,t)$ is reflecting particle--number conservation,
and
\ea
\label{chi-T-of-qt}
\chi_\perp(q,t)&=&\frac{1}{2}\sum_{\alpha,\alpha'}\left(\delta_{\alpha\alpha'}-\frac{q_\alpha
q_{\alpha'}}{q^2}\right)\chi_{\alpha\alpha'}({\bf q},t)\nonumber\\
&=&\frac{n}{\ii \hbar}\sum_{{\bf k}}C_{\bf k}\left(\frac{\hbar
k}{m}\right)^2 \left(1-\xi^2\right)\sin\left( \frac{\Delta_{{\bf
k}}({\bf q})}{\hbar}\,t\right)
\ee
with $\xi\g{\bf q}\cdot{\bf
k}/(qk)$.

Denoting by $\psi(q,t)$ any one of the response functions in Eqs.(\ref{FNIP-chi}--\ref{chi-T-of-qt}), the corresponding dynamical susceptibility $\tilde{\psi}(q,z)$  for $\Im z>0$ is defined as
\1
\label{def-dynamic-susc} \tilde{\psi}(q,z)=\ii\int_0^\infty\de
t~\e^{\ii t
z}\psi(q,t)=\int_{-\infty}^{\infty}\frac{\de\omega}{\pi}~\frac{\psi''(q,\omega)}{\omega-z}
\2
with the spectral function ($\omega$ real) given by
\1
\label{def-spectral-f}
\psi''(q,\omega)=\frac{1}{2}\int_{-\infty}^\infty\de t~\e^{\ii
t\omega}\psi(q,t)=\Im\lim_{\epsilon\to0}\tilde{\psi}(q,\omega+\ii
\epsilon)
\2
and for the related static susceptibility $\tilde{\psi}(q)$, one finds from
Eq.(\ref{def-dynamic-susc})
\1
\label{def-static-susc}
\tilde{\psi}(q)
=\lim_{\epsilon\to0}\lim_{\omega_0\to0}\tilde{\psi}(q,\omega_0+\ii
\epsilon)=\pp\int_{-\infty}^{\infty}\frac{\de\omega}{\pi}~\frac{\psi''(q,\omega)}{\omega}\;,
\2
where the Cauchy principal value is to be evaluated.

Applying Eqs. (\ref{def-dynamic-susc}) and (\ref{def-spectral-f}) on the results given in Eqs.(\ref{FNIP-chi}) and (\ref{FNIP-chi-alpha-beta}),  we obtain the dynamical susceptibilities \1
\label{FNIP-chiofz}
\tilde{\chi}(q,z)=2n\sum_{\bf k}C_{\bf k}\,f_1\left(\hbar
z,\,\Delta_{\bf k}({\bf q})\right),
\2
\1
\label{FNIP-chi-long-ofz}
\tilde{\chi}_\parallel(q,z)=\frac{n}{m}+\frac{z^2}{q^2}\,\tilde{\chi}(q,z),
\2
\1
\label{FNIP-chi-trans-ofz}
\tilde{\chi}_\perp(q,z)= \frac{n\hbar^2}{m^2}\sum_{\bf k}C_{\bf
k}\,k^2(1-\xi^2)\,f_1\left(\hbar z,\,\Delta_{\bf k}({\bf q})\right),
\2
and the  corresponding spectral functions
\ea
\label{chiimqom-1}
\chi''(q,\omega)&=&2n\sum_{\bf k}C_{\bf k}\,f_2\left(\hbar \omega,\,\Delta_{\bf k}({\bf q})\right),\\
\label{chilongim-1}
\chi_\parallel''(q,\omega)&=&\frac{\omega^2}{q^2}\,\chi''(q,\omega),
\ee
\1
\label{chitransim-1}
\chi_\perp''(q,\omega)= \frac{n\hbar^2}{m^2}\sum_{\bf k}C_{\bf
k}\,k^2(1-\xi^2)\,f_2\left(\hbar \omega,\,\Delta_{\bf k}({\bf
q})\right).
\2
The functions $f_1$ and $f_2$ appearing in the above
equations are
\1
\label{def-f1}
f_1(\varepsilon,\,\Delta)=\frac{\Delta}{\Delta^2-\varepsilon^2} \2
and \1 \label{def-f2} f_2(\varepsilon,\,\Delta)=\frac{\pi}{2}
\left[\delta(\varepsilon-\Delta)-\delta(\varepsilon+\Delta)\right].
\2

Another quantity of physical interest, closely related to the current spectra, is the normalized Kubo relaxation function which determines the evolution of a small sinusoidal initial current fluctuation
${\bf j}_{\parallel,\,\perp}({\bf r},t\g0)={\bf v}_{\parallel,\,\perp}\cos({\bf q}\cdot{\bf r})$ of the fluid according to
\1
\label{current-relax3} {\bf j}_{\parallel,\,\perp}({\bf r},t)=\phi_{\parallel,\,\perp}(q,t)~
{\bf v}_{\parallel,\,\perp}\cos({\bf q}\cdot{\bf r})\;.
\2
Here the spectral function $\phi_{\parallel,\perp}''(q,\omega)$ corresponding to $\phi_{\parallel,\,\perp}(q,t)$  is related to the current spectra by (see e.g. Ref.\cite[Chap. 3.5]{for:75})
\1
\label{current-relax2}
\phi_{\parallel,\perp}''(q,\omega)=
\frac{\chi_{\parallel,\perp}''(q,\omega)}{\omega}\left[
\pp\int_{-\infty}^\infty \frac{\de\bar{\omega}}{\pi}
\frac{\chi_{\parallel,\perp}''(q,\bar{\omega})}{\bar{\omega}}\right]^{-1}
\2
reflecting Kubo's identity.

The evaluation of the correlation functions requires knowledge of
$C_{\bf k}$  given by
\1
\label{concentration-Ck}
C_{\bf k}=\frac{g_s}{{\cal N}}\,\frac{\lambda}{\e^{\beta\varepsilon_k}-\eta\lambda}\;,
\2
where $g_s\g 2s+1$ is the spin-degeneracy factor for particle's spin $s$, $\lambda\g \e^{\beta \mu}$ is the fugacity, and $\eta = +1, -1, 0$ correspond respectively to a BE gas, an FD gas, and an MB gas.
The normalization condition
$\sum_{\bf k}C_{\bf k} = 1$ yields the implicit equation
\1
\label{mu-implicit}
\frac{g_s}{\eta}~\zeta_{3/2} \left(\eta\lambda\right)\g n\Lambda^3[1-C_0(T)]
\2
that determines the chemical potential $\mu \g \mu_{\eta}(n,T)$ of a uniform quantum gas. Here $\zeta_\nu(z)$ denotes the polylogarithm of order $\nu$, $\Lambda\g \left(2\pi\hbar^2\beta/m\right)^{1/2}$ is the thermal de Broglie wavelength and
\1
\label{cond-frac}
C_0(T)=\delta_{\eta,\,1}\,\Theta\left(T_c-T\right)\left[1-\left( T/T_c\right)^{3/2}\right],
\2
is the average fraction of particles occupying the ground state, wherein $\delta_{i,\,j}$ denotes the Kronecker delta, $\Theta(x)$ the unit step function, and $T_c\g\left(2\pi\hbar^2/mk_B\right)\left(n/g_s \zeta(3/2)\right)^{2/3}$ the Bose condensation temperature;  $\zeta(3/2)$ is the Riemann zeta function.

\section{Spectral functions}

\label{current-spectra}

The spectral functions  may be calculated explicitly by making use
of ${\bf k}$-sum expressions derived in Appendix \ref{AppSum}.
Choosing $f(\varepsilon)\g2n \, f_2(\hbar\omega,\varepsilon)$ in
Eq.(\ref{sumS1-3}) and \\$f(\varepsilon)\g  n\hbar^2/m^2 \,
f_2(\hbar\omega,\varepsilon)$ which implies $F(\varepsilon)\g n\pi\hbar^2/(2m^2)[\Theta(\epsilon-\hbar\omega)-\Theta(\epsilon+\hbar\omega)]$ for the indefinite integral in Eq.(\ref{sumS2-3}), the ${\bf
k}$--sums in Eqs.(\ref{chiimqom-1}) and (\ref{chitransim-1}) are
evaluated to give for the density response spectrum
\1
\label{chiimqom-2}
\chi''(q,\omega)=2C_0(T)
\,n\pi\,\hbar\omega\,\delta\left(\hbar^2\omega^2-\varepsilon_q^2\right)
+A~G_1(x,b,\eta\lambda)\;,
\2
and for the transverse--current response spectrum
\1
\label{chitransim-2}
\chi''_\perp(q,\omega)=\frac{A}{m\beta}\,G_2(x,b,\eta\lambda)
\2
with
$x\g \beta\hbar\omega/\left(2\sqrt{\beta\varepsilon_q}\right)$,
$b\g\sqrt{\beta\varepsilon_q}/2 = q\Lambda/\left(4\sqrt\pi\right)$,
\1
\label{abbrev-F}
G_\nu(u,v,w)=\frac{1}{v\,w}\left[\zeta_\nu\left(\e^{-(u-v)^2}w\right)-
\zeta_\nu\left(\e^{-(u+v)^2}w\right)\right]
\2
and
\ea
\label{pre-factor-A} A&=&\frac{g_s\lambda\,\beta\sqrt{\pi}}{4\Lambda^3}
=\frac{[1-C_0(T)]\,\eta\lambda\,\sqrt{\pi}}{4\,\zeta_{3/2}\left(\eta\lambda\right)}\,n\beta\;,
\ee
wherein the rewriting of $A$ follows from Eq.(\ref{mu-implicit}), and the recursion relation $\zeta_{\nu+1}(u)=\int_0^u\de z~z^{-1}\zeta_{\nu}(z)$ with $\zeta_{0}(z)=z/(1-z)$, see, e.g. Ref. \cite{lee:95}, has been used.

\begin{figure}\begin{center}
\includegraphics[width=80mm,angle=0]{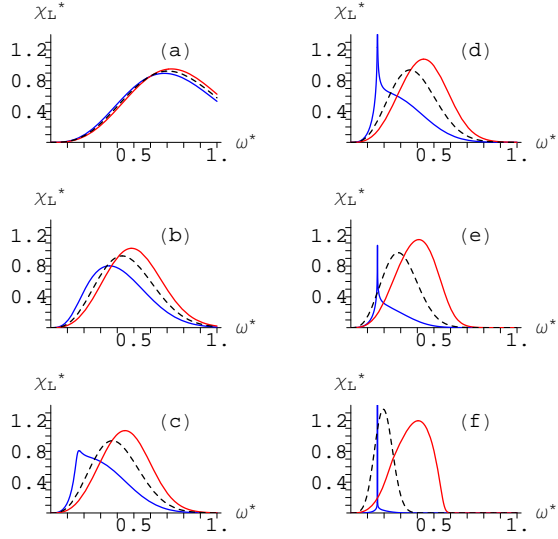}
\parbox{120mm}{\caption{\label{FNIP-chi-long}\small
(Color online) Longitudinal--current response spectra of uniform
ideal quantum gases, $\chi_{\rm
L}^*(q^*,\omega^*;T^*)\g(2m/n\pi)\,\chi_{\parallel}''(q,\omega)$ for $q^*\g0.4$ as a function of $\omega^*$: BE gas (blue), FD gas (red), and MB gas (dashed) at reduced temperatures $T^*\g0.491$(a), 0.164(b),  0.114(c),  0.104(d),  0.055(e),  0.011(f) corresponding to
$T/T_c\g4.5$, 1.5, 1.05, 0.95, 0.5, 0.1, respectively.
}}\end{center}
\end{figure}

Noting that $G_\nu(x,b,\eta\lambda)\stackrel{\eta\to0}{\longrightarrow}
\left(1/b\right)\left(1-\e^{-4bx}\right)\e^{-(x-b)^2}$,
one recovers from Eqs.(\ref{chiimqom-2}) and (\ref{chitransim-2}) the known results for an ideal Boltzmann fluid
($v_T\g1/\sqrt{\beta m}$):
\ea
\label{MB-response1}
\chi''(q,\omega)&\stackrel{\eta\to0}{\longrightarrow}&\frac{n\sqrt{\pi}}{\sqrt2\,\hbar
qv_T}\left(1-e^{-\beta\hbar\omega}\right) \exp\left[-
\frac{(\hbar\omega-\varepsilon_q)^2}{2v_T^2\hbar^2 q^2}\right]\;, \\
\label{MB-response2}
\chi_\perp''(q,\omega)&\stackrel{\eta\to0}{\longrightarrow}&v_T^2\chi''(q,\omega)\;,
\ee
which will reduce to the corresponding expressions for
a {\em classical} gas \cite{ham:05}, if the limit $\hbar\to0$ is performed.

For plotting purposes, we introduce the dimensionless temperature $T^*\g k_{\rm
B}T/\varepsilon_{\rm u}$, where $\varepsilon_{\rm u}\g\hbar^2k_{\rm u}^2/2m$ and $k_{\rm u}\g 2\left(6\pi^2n/g_s\right)^{1/3}$
denote system--dependent units of energy and wavenumber, respectively. We have $T^*_c=\left[6\sqrt\pi \zeta(3/2)\right]^{-2/3}\approx0.109$  and, in the case of fermions with $k_{\rm F}$ as the Fermi wavenumber, $k_{\rm u}\equiv 2k_{\rm F}$.
We also introduce dimensionless {\em reduced} spectral functions, $
\chi^*_{\rm L,T}(q^*,\omega^*;T^*)\g(2m/n\pi)\,\chi_{\parallel,\perp}''(q,\omega)
$, which depend only on the dimensionless parameters $q^*\g q/k_{\rm u}$, $\omega^*\g\hbar\omega/\varepsilon_{\rm u}$ and $T^*$.
\begin{figure}\begin{center}
\includegraphics[width=80mm,angle=0]{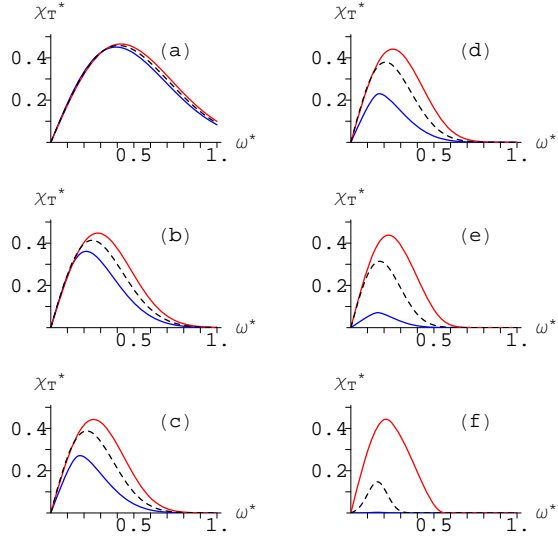}
\parbox{120mm}{\caption{\label{FNIP-chi-trans}\small
(Color online) Transverse--current response spectra of uniform ideal
quantum gases, $\chi_{\rm T}^*(q^*,\omega^*;T^*)\g(2m/n\pi)\,\chi_\perp''(q,\omega)$. Parameters as in
Fig.\ref{FNIP-chi-long}. }}\end{center}
\end{figure}

The reduced spectra, which follow from Eqs.(\ref{FNIP-chi-long-ofz}), (\ref{chiimqom-2}) and (\ref{chitransim-2}) for uniform ideal quantum gases, are displayed in Figs.\ref{FNIP-chi-long} and \ref{FNIP-chi-trans} as
a function of $\omega^*$ for one fixed $q^*\g0.4$. Current response spectra are {\em odd} functions of frequency and hence  they are displayed for $\omega^*\ge0$ only.
The graphs (a)--(f)
in each figure refer to spectra calculated for decreasing
temperatures ranging from high ($T\gg T_c$) to low ($T\ll T_c$)
values. It can be seen that due to $T$ being sufficiently high in
Figs.\ref{FNIP-chi-long}(a) and
\ref{FNIP-chi-trans}(a), there is almost no difference between the
bosonic and the fermionic spectral functions and, as expected, both nearly agree
with the response spectrum obtained for the MB gas.
In Figs.\ref{FNIP-chi-long}(b) and (c), the
Bose gas spectrum is seen to develop a precursor to the sharp
free--boson excitation peak at $\omega^*\g q^{*2}\g0.16$, although the
temperatures are well above $T_c\,$. The spectral weight in the longitudinal boson--current spectra appears deformed due to {\em thermal} excitations which, owing to the $\omega^2$--factor in the expression corresponding to Eq.(\ref{FNIP-chi-long-ofz}), get amplified more for larger $\omega$. This mechanism, which is not effective in the case of the sharp $\delta$--function resonance below $T_c$ reflecting boson excitations out of the condensate, gives rise to the peculiar separation of spectral weight visible in Figs.\ref{FNIP-chi-long}(c)--(e).

As opposed to the longitudinal--current response, there is no contribution from the boson condensate to the transverse--current response (see Fig.\ref{FNIP-chi-trans}). This is due to the factor $k^2$ in Eq.(\ref{FNIP-chi-trans-ofz}) which suppresses the (${\bf k}\g0$)--term in the sum and leaves only contributions of thermally excited bosons with ${\bf p}\g\hbar{\bf k}\ne0$. Hence, similar to the thermal contributions in case of the longitudinal--current response (Fig.\ref{FNIP-chi-long}), the transverse--current response spectrum will vanish completely as $T\to0$ (Fig.\ref{FNIP-chi-trans}), because all bosons condense into the zero--momentum state. The maximum value of
$\chi''_\perp(q,\omega)$, which is reached at
$\omega\g\varepsilon_q/\hbar$ for low temperatures, can be seen to vanish as
$T^2$ when temperature is decreased to zero. The interesting
behavior found for the transverse--current response of the
uniform ideal Bose gas has a noteworthy implication to be discussed in  Sec. \ref{Condensation and Superfluidity}.

\begin{figure}\begin{center}
\includegraphics[width=80mm,angle=0]{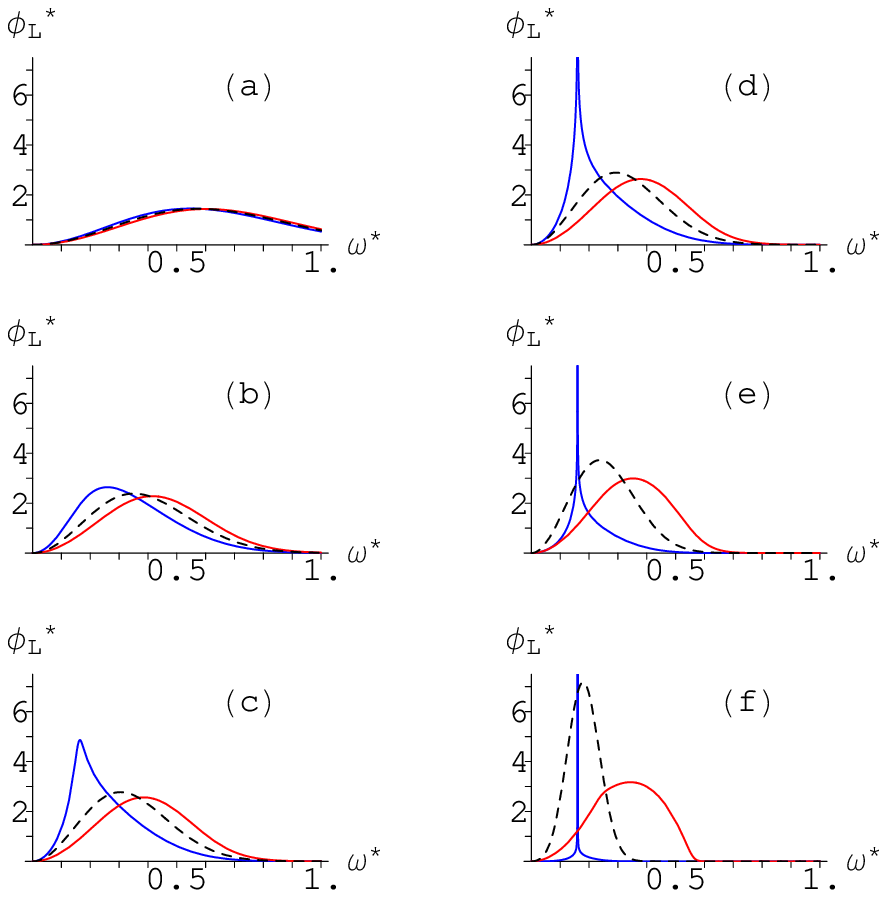}
\parbox{120mm}{\caption{\label{FNIP-normphi-long}\small
(Color online) Reduced normalized longitudinal--current relaxation
spectra of uniform ideal quantum gases, $\phi^*_{\rm
L}(q^*,\omega^*;T^*)\g2\varepsilon_{\rm u}/(\hbar\pi)\,\phi_\parallel''(q,\omega).$
Parameters as in Fig.\ref{FNIP-chi-long}. }}\end{center}
\end{figure}

\begin{figure}\begin{center}
\includegraphics[width=80mm,angle=0]{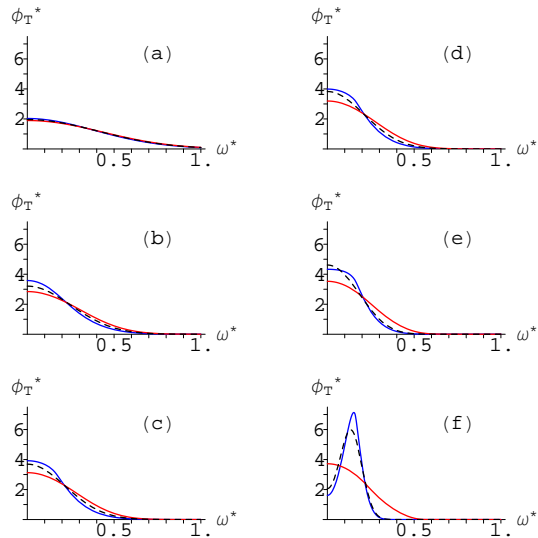}
\parbox{120mm}{\caption{\label{FNIP-normphi-trans}\small
(Color online) Reduced normalized transverse--current relaxation
spectra of uniform ideal quantum gases, $\phi^*_{\rm
T}(q^*,\omega^*;T^*)\g2\varepsilon_{\rm u}/(\hbar\pi)\,\phi_\perp''(q,\omega).$
Parameters as in Fig.\ref{FNIP-chi-long}. }}\end{center}
\end{figure}

We also calculated the normalized current relaxation spectra of longitudinal and transverse current fluctuations according to Eq.(\ref{current-relax2}). In Figs.\ref{FNIP-normphi-long} and \ref{FNIP-normphi-trans}, reduced spectra
$
\phi^*_{\rm L,T}(q^*,\omega^*;T^*)=2\varepsilon_{\rm u}/(\hbar \pi)\,\phi_{\parallel,\perp}''(q,\omega)
%=\frac{\chi^*_{\rm L,T}(q^*,\omega^*;T^*)/\omega^*}{\int_0^\infty\de\bar{\omega}^* ~\chi^*_{\rm L,T}(q^*,\bar{\omega}^*;T^*)/\bar{\omega}^* }
$
are displayed. As opposed to the response spectra, the relaxation spectra are {\em even} functions of $\omega^*$, which are normalized according to $\int_0^\infty\de\omega^*~\phi^*_{\rm L,T}(q^*,\omega^*;T^*)\g 1$.

The longitudinal spectrum $\phi_{\rm L}^*$ for a BE gas (blue lines in
Fig.\ref{FNIP-normphi-long}) develops a sharp peak, as the
temperature is approaching $T_c$ from above, with an additional $\delta$--function peak for $T< T_c$. This peak can here be traced back to the same excitations of condensed bosons which are also responsible for the corresponding peaks in Fig.\ref{FNIP-chi-long}.  According to Eq.(\ref{current-relax3}), these sharp peaks have the following physical implications. Any small longitudinal current fluctuation  of large wavelength will perform undamped oscillations of very low frequency, i.e. the flow will sustain for long times. Near the surface of a fluid container, spontaneous current fluctuations across the surface will be of {\em longitudinal} nature resulting in the undamped creeping flow across the container surface as observed in superfluids. This relaxation behavior of longitudinal current fluctuations in a uniform ideal Bose gas below $T_c$ fits in with the implications of the transverse current response discussed in  Sec. \ref{Condensation and Superfluidity}.

Contrary to the longitudinal current relaxation, the transverse counterpart $\phi_{\rm T}^*$
behaves surprisingly ``normal''. In fact, the transverse--current relaxation spectrum of a Bose gas is very similar, at all temperatures, to that of the Boltzmann
gas (cf. Fig.\ref{FNIP-normphi-trans}), since the condensate does not contribute as mentioned earlier in this section. It is noteworthy to point
out that at low temperatures the transverse relaxation spectra of BE and MB
gases, as opposed to the FD gas, show a maximum at the same non--zero frequency, $\omega_{\rm max}\approx\epsilon_q/\hbar$ (cf. Eq.(\ref{MB-response2})).

\section{Temperature--Dependent Static Susceptibilities}
\label{Static Susceptibilities}

We insert Eq.(\ref{chilongim-1}) together with Eq.(\ref{chiimqom-2}) into Eq.(\ref{def-static-susc}) and then use $\int_{-\infty}^\infty~\de x \,\zeta_1(\alpha e^{-x^2})=\sqrt{\pi}~\zeta_{3/2}(\alpha)$ to demonstrate that
\1
\label{f-sum-rule}
\tilde{\chi}_{\parallel}(q)=\frac{n}{m}
\2
for all $q$ and $T$, which reflects the famous $f$--sum rule (and which could be derived more easily by taking $z\to 0$ in Eq.(\ref{FNIP-chi-long-ofz}), of course).
However, the evaluations of $\tilde\chi{(q)}$ and $\tilde{\chi}_{\perp}(q)$ are more tedious.
The use of Eq.(\ref{def-static-susc}) in conjunction with Eq. (\ref{chiimqom-2}) and Eq. (\ref{chitransim-2}), respectively, gives
\1
\label{ChiTilde-q-1}
\tilde{\chi}(q)=n\beta\left[\frac{C_0(T)}{2b^2}+ \frac{2A}{n\beta\pi}\pp\int_{0}^\infty\frac{\de x}{x}\,G_1\left(x,b,\eta\lambda\right)\right]
\2 and
\1
\label{ChiTildePerp-q-1}
\tilde{\chi}_\perp(q)=\frac{2A}{\pi m\beta}\pp\int_{0}^\infty\frac{\de x}{x}\,G_2\left(x,b,\eta\lambda\right)\;.
\2

For $\lambda<1$, i.e. for {\em negative} chemical potential of the gas, the above integrals may be evaluated exactly as a power series involving error functions with imaginary arguments which we finally express as
\ea
\label{ChiTilde-q-2}
\tilde{\chi}(q)&=&n\beta\left[\frac{8\pi C_0(T)}{\Lambda ^2 q^2}+ \frac{[1-C_0(T)]}{\zeta_{\frac{3}{2}}\left(\eta\lambda\right)}\sum_{\ell=1}^\infty \frac{\left(\eta\lambda\right)^\ell}{\ell^{\frac{1}{2}}}\, {_1}F_1\left(1;\frac{3}{2};-\ell b^2\right)\right]\\
\label{asympChiTilde-q-2}
&\stackrel{q\to0}{\longrightarrow}& n\beta\left[\frac{8\pi C_0(T)}{\Lambda ^2 q^2}\right.\nonumber\\
&&\left.\hspace{3em}+ [1-C_0(T)]\, \left\{\frac{\zeta_{\frac{1}{2}}(\eta\lambda)}{\zeta_{\frac{3}{2}}(\eta\lambda)} -\frac{\Lambda^2}{24\pi}\frac{\zeta_{-\frac{1}{2}}\left(\eta\lambda\right)}{\zeta_{\frac{3}{2}}(\eta\lambda)}\, q^2+{\cal O}(q^4)\right\}\right]
\ee
and
\ea
\label{ChiTildePerp-q-2}
\tilde{\chi}_\perp(q)&=&\frac{n[1-C_0(T)]}{m\zeta_{\frac{3}{2}}\left(\eta\lambda\right)}\sum_{\ell=1}^\infty \frac{\left(\eta\lambda\right)^\ell}{\ell^{\frac{3}{2}}}\, {_1}F_1\left(1;\frac{3}{2};-\ell b^2\right)\\
\label{asympChiTildePerp-q-2}
&\stackrel{q\to0}{\longrightarrow}& \frac{n}{m}[1-C_0(T)]\left\{1
-\frac{\Lambda^2}{24\pi}\frac{\zeta_{\frac{1}{2}}\left(\eta\lambda\right)}{\zeta_{\frac{3}{2}}\left(\eta\lambda\right)}\,
\;q^2+{\cal O}(q^4)
\right\}
\ee
with ${_1}F_1\left(\alpha_1;\alpha_2;z\right)$ as the Kummer function, one of the confluent hypergeometric functions \cite{abs:72}.
For an MB gas ($\eta\g0$), only the first term ($\ell=1$) of the series survives leading to closed analytical forms for both $\tilde{\chi}(q)$ and $\tilde{\chi}_\perp(q)$.

Whereas for a Bose gas  the above expressions are valid at all temperatures, for a Fermi gas numerical evaluation of Eqs. (\ref{ChiTilde-q-1}) and (\ref{ChiTildePerp-q-1}) is required in the regime $\lambda>1$, in general. We note, however, that the asymptotes given in Eqs. (\ref{asympChiTilde-q-2}) and (\ref{asympChiTildePerp-q-2}) reproduce the small--$q$ behavior of the susceptibilities  correctly not only for BE and MB gases but also for a FD gas at {\em all} temperatures.
The asymptotic Eqs.(\ref{asympChiTilde-q-2}) and (\ref{asympChiTildePerp-q-2}) were deduced with the help of the Kummer--function series expansion ${_1}F_1\left(1;3/2;-\ell b^2\right)=\sum_{p\g0}^\infty \left(-\ell\Lambda^2 q^2 /8\pi\right)^p /\left(2p+1\right)!!$. But a word of caution needs to be in place regarding any temptation to use the latter in combination with a subsequent change of order of summations in Eqs. (\ref{ChiTilde-q-2}) and (\ref{ChiTildePerp-q-2}); such a procedure would result in a divergent power series in $q$.

\begin{figure}
\begin{center}
\includegraphics[width=90mm,angle=0]{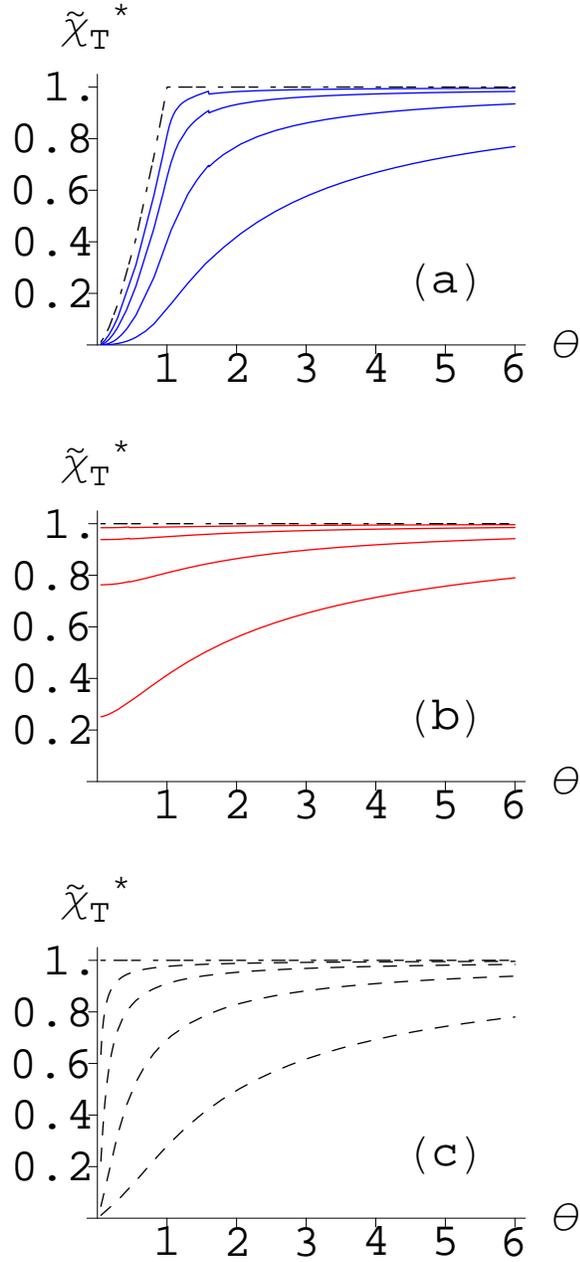}
\parbox{120mm}{\caption{\label{FNIP-chiTofq}\small
(Color online) Reduced transverse--current susceptibility
$\tilde{\chi}^*_{\rm T}\g\tilde{\chi}_\perp(q)m/n$ as a function of the
temperature ratio $\Theta\g T/T_c$ for uniform ideal quantum gases
obeying BE (a), FD (b), and MB (c) statistics. In each
plot, the curves  refer to wavenumbers $q^*\g 0.0,\,0.125,\,0.25,\,0.5,\,1.0$, respectively, from the top dash--dotted line to the bottom line. }}
\end{center}
\end{figure}

It is seen that contrary to $\tilde{\chi}_{\parallel}(q)$ in Eq. (\ref{f-sum-rule}), $\tilde{\chi}_{\perp}(q)$ depends on both $q$ and $T$. Figure \ref{FNIP-chiTofq} displays the ratio $\tilde{\chi}^*_{\rm
T}\g\tilde{\chi}_{\perp}(q)/\tilde{\chi}_{\parallel}(q)$ as a function
of $\Theta\g T/T_{\rm c}$ for a set of wavenumbers in the range $0\le q^*\le1$ for all the three gases, and with
$T_c$ as the value corresponding to a uniform ideal Bose gas.   Figures \ref{FNIP-chiTofq}(a) and (c) show that the BE and MB gases have rather similar behavior  for $T\to 0$ except that the former shows a thermodynamic phase transition at a finite temperature. The values $\tilde{\chi}_{\perp}(q)\ne0$ for a Fermi gas at $T=0$ in Fig. \ref{FNIP-chiTofq} (b) can be traced back to the fact that all states fill completely up to the Fermi energy.  As T increases, $\tilde{\chi}_{\perp}(q\ne0)$ increases and asymptotically approaches $\tilde{\chi}_{\parallel}(q)=n/m$ irrespective of the statistics signifying that each of the gases considered here responds identically to transverse and longitudinal probes at high temperatures.

Since $\lim_{v\to0}G_{\nu}(u,v,w)\g4 u \,\zeta_{\nu-1}\left(w \,\e^{-u^2}\right)/w$, one finds from Eqs.(\ref{f-sum-rule}) and (\ref{ChiTildePerp-q-1}) that
uniform ideal FD and MB gases at all temperatures, and a BE gas
at $T\ge T_c$, will obey
\1
\label{normal-fluid}
\lim_{q\to0}\tilde{\chi}_{\perp}(q)=\lim_{q\to0}\tilde{\chi}_{\parallel}(q)=\frac{n}{m}\;.
\2
 On the other hand, for bosons at $T<T_c\,$ we have
\1
\label{super-fluid}
\lim_{q\to0}\tilde{\chi}_{\perp}(q)=\frac{n}{m}[1-C_0(T)]<\lim_{q\to0}\tilde{\chi}_{\parallel}(q)\;,
\2
with the finite condensed fraction, $C_0(T)>0$ given by Eq.(\ref{cond-frac}). The inequality in Eq. (\ref{super-fluid}) achieved through explicit evaluation of transverse and longitudinal particle--current susceptibilities generalizes the $T\g0$ result mentioned in Ref.\cite[Chap. 4, Vol.II]{pin:99}.

\section{Condensation and Superfluidity}
\label{Condensation and Superfluidity}

This section constitutes a spin--off of our investigations in earlier sections. We wish to apply Martin's criterion to uniform quantum gases and closely follow the elaborations given in \cite{for:75}. The $f$--sum rule given by Eq. (\ref{f-sum-rule}) is a direct consequence of the gauge invariance and hence it holds for any fluid. The total mass density $\rho=mn$ of a one--component fluid, interacting or noninteracting, is therefore defined as
\1
\label{total density}
\rho=m^2\tilde{\chi}_\parallel(q\to0)=m^2\lim_{q\to0}\pp\int_{-\infty}^\infty\frac{\de\omega}{\pi}~\frac{\chi_\parallel''(q,\omega)}{\omega}.
\2
One defines another mass density $\rho_n$ by the relation
\1
\label{normal density}
\rho_n=m^2\tilde{\chi}_\perp(q\to0)=m^2\lim_{q\to0}\pp\int_{-\infty}^\infty\frac{\de\omega}{\pi}~\frac{\chi_\perp''(q,\omega)}{\omega}
\2
and one finds $\rho_n=\rho$ from Eq.(\ref{normal-fluid}),  implying momentum correlations of {\em finite} range which is characteristic of a {\em normal} fluid. However, the result from Eq.(\ref{super-fluid}), $\rho_n<\rho$, implies that the gas has momentum correlations of infinite range which is a characteristic of a superfluid \cite [Sec. V.E]{hom:65},\cite[Chap. 10]{for:75}.
Thus the difference $\rho_s\g\rho-\rho_n$, known as superfluid density, turns out to be equal to the condensate density $\rho C_0(T)$, a result also reported in Ref.\cite{fbj:73}.

We provide an independent consistency check for the expression $\rho_s=\rho\, C_0(T)$
using the Josephson relation \cite{jos:66} rewritten
in the form
\1
\label{josephson}
\pp\int_{-\infty}^\infty\frac{\de\omega}{2\pi\hbar}~\frac{G''({\bf
k},\omega)}{\omega}=\frac{(m/\hbar)^2}{\rho_s k^2}~n_0\;,
\2
which relates the superfluid (mass) density $\rho_s$ to the condensate (number) density
$n_0\g n\, C_0(T)$. Here $G''({\bf k},\omega)$ denotes the spectral
function
\1
G({\bf r}-{\bf r'},t-t'):=\frac{1}{g_s}\sum_\sigma\erw{\left[a_{\sigma}({\bf
r},t)\,,~ a_{\sigma}^\dagger({\bf r'},t')\right]}
\2
corresponding to the correlation function of boson
field operators. For a uniform ideal Bose gas, one finds $ G''({\bf k},\omega)\g \pi \delta(\omega-\varepsilon_k/\hbar)$  and hence obtains from Eq. (\ref{josephson}),  $\rho_s\g mn_0$ in agreement with our finding.

We note that the existence of long--range correlations together with our discussions in Sec. \ref{current-spectra} of long--wavelength longitudinal--current relaxation spectra  lead to the conclusion that a uniform ideal Bose gas at $T<T_c$ possesses some characteristic properties of a superfluid.
However, there is a well--known result, which can also be retrieved from Eq. (\ref{ChiTilde-q-2}), that the static compressibility \cite{kappa:09} given by $\kappa\g\lim_{q\to0}\kappa(q)\g\lim_{q\to0}\left[\tilde{\chi}(q)/n^2\right]$  diverges for $T<T_c$ for an ideal Bose gas implying a sound velocity $c\g0$ in the condensed phase. Hence such a system does not satisfy Landau's requirement for superfluidity, namely that of a nonzero critical velocity (see e.g. \cite {pis:03,bdz:08}).
This in turn suggests that the existence of long--range momentum correlations will only represent a necessary condition for a system to be superfluid.
On the other hand, an interacting Bose gas, howsoever weak the interaction might be, has finite compressibility and sound velocity, and would be capable of sustaining a superflow.

We, therefore, consider now particles interacting repulsively via a {\em short--range} pair potential $v(|{\bf r}|)$ whose influence can be incorporated approximately on our findings of previous sections by applying the RPA. The expressions for the dynamical susceptibilities given by Eqs. (\ref {FNIP-chiofz}) to (\ref {FNIP-chi-trans-ofz}) then take the forms
\ea
\label{RPA-D}
\tilde{\chi}^{\rm RPA}(q,z)&=&\frac{\tilde{\chi}(q,z)}{1+\tilde{v}({\bf q})\,\tilde{\chi}(q,z)}\;,\\
\label{RPA-LC}
\tilde{\chi}^{\rm RPA}_\parallel(q,z)&=&\frac{n}{m}+\frac{z^2}{q^2}\,\tilde{\chi}^{\rm RPA}(q,z)\;,\\
\label{RPA-TC}
\tilde{\chi}^{\rm RPA}_\perp(q,z)&=&\tilde{\chi}_\perp(q,z)\;.
\ee
While Eq.(\ref{RPA-D}) is a well known RPA result \cite [vol. I]{pin:99} and Eq.(\ref{RPA-LC}) follows from it as an immediate consequence of number conservation (cf. Eq. (\ref {FNIP-chi-long-ofz})), Eq.(\ref{RPA-TC}) has been deduced from Eq. (12) of Ref. \cite{sip:75}. We note that Eq. (\ref{RPA-TC}) can serve as an approximate expression for the transverse particle--current susceptibility of an  interacting system of {\em neutral} quantum particles only. For {\em charged} particles, transverse electromagnetic shielding effects will modify Eq.(\ref{RPA-TC}) due to  intimate coupling between particle current and charge current, as discussed in Paper II \cite{bps.b:09}. It may be remarked that the considered approximation is a consistent weak--coupling theory having its validity in the limiting situation $\tilde{v}({\bf q})\to0$ for short--range interactions.

Assuming the pair potential to be a contact interaction,  one has $\tilde{v}({\bf q})\g4\pi\hbar^2 a/m$, with $a>0$ denoting the scattering length. The compressibility of the interacting Bose gas turns out to be finite in the condensed phase, which is readily seen by taking the static limit in Eq.(\ref{RPA-D}) and using $\tilde{\chi}(q)$ from Eq.(\ref{ChiTilde-q-1}). This removes the pathological divergence of the compressibility of the non--interacting gas and results in a finite sound velocity $c$,  a necessary condition for the superfluidity  of the Bose--condensed phase.
It follows from Eq. (\ref{RPA-LC}) that although RPA modifies the longitudinal dynamical  susceptibility,  its  {\em static} value  remains unaltered, as expected. Furthermore,  Eq. (\ref{RPA-TC}) implies that, within the RPA, interactions do not alter the dynamic, and hence the static, transverse--current susceptibility. This, therefore, implies that the relation between superfluid density  and condensate density remains the same as in the ideal Bose gas.

\section{Summary}
\label{Conclusion}

We have presented a comprehensive study of the dynamics of uniform ideal quantum gases in terms of the number--density and the particle--current density response functions. Unified expressions for the associated response spectra, valid at any wavenumber $q$, frequency $\omega$, and temperature $T$, as well as the corresponding static susceptibilities have been obtained for BE, FD and  MB gases. All results, many of them given in {\em exact analytical} forms and believed to be new, have been discussed in view of their applications in various contexts. The plots of the response spectra provide an immediate comparison of the dynamical behaviour, indicating intrinsic and apparent differences due to the statistics.
The {\em relaxation} spectra of longitudinal particle--current fluctuations have been found to develop a peak as  $T$ is approaching $T_c$ from above and lowered below $T_c$, contrary to their transverse counterparts. The transverse--current relaxation spectra of the BE gas are surprisingly similar to those of an MB gas for all $T$.

It is found in  the long--wavelength limit that the static transverse boson--current susceptibility approaches a $T$--dependent value equal to $\rho_n/m^2<\rho/m^2$  below $T_c$. From this result we conclude that  in a condensed ideal Bose gas the momentum correlations are of long range and the superfluid density equals the condensate density.
Repulsive contact interactions have been taken into account within the RPA. This approximation removes the divergence of the compressibility and leads to a transverse particle--current susceptibility equal to that of an ideal gas. Our studies  demonstrate that the existence of long--range momentum correlations constitutes only a necessary condition for a system to be a superfluid.

\section*{Acknowledgments}

The work is partially supported by the Indo--German (DST--DFG) collaborative research program. JB and KNP gratefully acknowledge financial support from the Alexander von
Humboldt Foundation.

\section*{Appendix}

\renewcommand{\theequation}{\mbox{\Alph{section}.\arabic{equation}}}
\begin{appendix}

\setcounter{equation}{0}
\section{Evaluation of {$\bf k$}-Sums}
\label{AppSum}

We have to evaluate the {$\bf k$}-sums appearing in
Sec. \ref{basics} of the forms $S\g\sum_{\bf
k}C_{\bf k}\, f(\Delta_{\bf k}({\bf q}))$ and $S_\perp\g\sum_{\bf
k}C_{\bf k}\,k^2(1-\xi^2) f(\Delta_{\bf k}({\bf q}))$ . Separating out the term $\bf k\g0$,
converting the remaining sum over $\bf k\!\ne\!0$ into the $\bf
k$-integration corresponding to a macroscopic system, and transforming to integration
variable $x\g\sqrt{\beta\varepsilon_k}$, we obtain
\ea
\label{sumS1-1}
S= C_0(T)\, f(\varepsilon_q)+\frac{g_s\lambda}{n\Lambda^3\sqrt{\pi}} \int_{-\infty}^\infty\de x~
\frac{x^2}{\e^{x^2}-\eta\lambda}\int_{-1}^{1}\de\xi~
f\left(\varepsilon_q+2\xi x\sqrt{\varepsilon_q/\beta}\right)\;,
\ee
and
\ea
\label{sumS2-1}
S_\perp=\frac{4\sqrt{\pi}\,g_s\lambda}{n\Lambda^5 }
\int_{-\infty}^\infty \de x~\frac{x^4}{\e^{x^2}-\eta\lambda}
\int_{-1}^{1}\de\xi~(1-\xi^2)f\left(\varepsilon_q+2\xi x\sqrt{\;\varepsilon_q/\beta}\right)\;.
\ee
For later convenience, $x$--integrations have been extended over the complete real axis since both integrands are even functions of $x$.
Using the identity
\ea
\label{identity}
\frac{1}{\e^{x^2}-\eta\lambda}=\frac{1}{2\eta\lambda\, x}\,\dnd{x}{}\ln\left(1-\eta\lambda\e^{-x^2}\right)\;,
\ee
the $x$-integrations are done by parts {\em before} evaluation of the $\xi$--integrals is attempted. This allows Eqs.(\ref{sumS1-1}) and (\ref{sumS2-1}) to be rewritten finally as one--dimensional integrals,
\ea
\label{sumS1-3}
S= C_0(T)\, f(\varepsilon_q)- \frac{g_s}{n\eta\sqrt{\pi}\Lambda^3}
\int_{-\infty}^\infty \de x~\ln\left(1-\eta\lambda e^{-x^2}\right)
f\left(\varepsilon_q+2 x\sqrt{\varepsilon_q/\beta}\right)
\ee
and
\ea
\label{sumS2-3}
S_\perp=-\frac{8g_s\pi\beta}{n\eta q\Lambda^6}\int_{-\infty}^\infty \de
x~x\ln\left(1-\eta\lambda\e^{-x^2}\right)
F\left(\varepsilon_q+2x \sqrt{\varepsilon_q/\beta}\right)\;,
\ee
where $F(x)\g\int \de x~f(x)$ is the indefinite integral of the function $f$.

\end{appendix}

\bibliographystyle{unsrt}

\begin{thebibliography}{1}
\bibitem{kappa:09}
It is noted that the static compressibility $\kappa$ for an ideal quantum gas
  coincides with the isothermal compressibility
  $\kappa_T\g-\frac{1}{V}\left(\pd{P}{V}\right)_{T}$. The latter is well known
  to diverge for an ideal Bose gas below $T_c$.
\end{thebibliography}

\end{document}